\documentstyle[10pt,epsfig,dp_delphititle,float,hangcaption,xspace,amssymb,
amsfonts,amsmath,amsthm,cite,graphicx]{dp_delphi}
\makeindex
\def\DpPaperGroup{PH-EP}
\def\DpPaperRef{2008--013}
\def\DpDate{29 August 2008}
\def\DpAuthors{DELPHI Collaboration}
\def\DpSubmit{(Accepted by Eur. Phys. J. C)}
\def\DpTitle{{ 
Search for one large extra dimension with the DELPHI detector at LEP
}}
\def\DpComment{}
\def\DpEMail{}

\newcommand {\eeGg} {$e^+e^- \rightarrow G\gamma$ \hspace*{1mm}}
\newcommand {\eenngg} {$e^+e^- \rightarrow \nu\bar{\nu}\gamma(\gamma)$}
\newcommand {\eeeeg} {$e^+e^- \rightarrow e^+e^-\gamma$ \hspace*{1mm}}
\begin{document}
\makeatletter
\newcount\@tempcntc
\def\@citex[#1]#2{\if@filesw\immediate\write\@auxout{\string\citation{#2}}\fi
  \@tempcnta\z@\@tempcntb\m@ne\def\@citea{}\@cite{\@for\@citeb:=#2\do
    {\@ifundefined
       {b@\@citeb}{\@citeo\@tempcntb\m@ne\@citea\def\@citea{,}{\bf ?}\@warning
       {Citation `\@citeb' on page \thepage \space undefined}}%
    {\setbox\z@\hbox{\global\@tempcntc0\csname b@\@citeb\endcsname\relax}%
     \ifnum\@tempcntc=\z@ \@citeo\@tempcntb\m@ne
       \@citea\def\@citea{,}\hbox{\csname b@\@citeb\endcsname}%
     \else
      \advance\@tempcntb\@ne
      \ifnum\@tempcntb=\@tempcntc
      \else\advance\@tempcntb\m@ne\@citeo
      \@tempcnta\@tempcntc\@tempcntb\@tempcntc\fi\fi}}\@citeo}{#1}}
\def\@citeo{\ifnum\@tempcnta>\@tempcntb\else\@citea\def\@citea{,}%
  \ifnum\@tempcnta=\@tempcntb\the\@tempcnta\else
   {\advance\@tempcnta\@ne\ifnum\@tempcnta=\@tempcntb \else \def\@citea{--}\fi
    \advance\@tempcnta\m@ne\the\@tempcnta\@citea\the\@tempcntb}\fi\fi}
 
\makeatother

\begin{titlepage}
\pagenumbering{roman}

\CERNpreprint{\DpPaperGroup}{\DpPaperRef}   
\date{{\small\DpDate}}                      
\title{\DpTitle}                            
\address{\DpAuthors}                        

\begin{shortabs}                            
\noindent
Single photons detected by the DELPHI experiment at LEP2 in the years
1997-2000 are reanalysed to investigate the existence of a single extra
dimension in a modified ADD scenario with slightly warped large extra dimensions.
The data collected at centre-of-mass energies between 180 and 209 GeV
for an integrated luminosity of $\sim$ 650 $\mathrm {pb^{-1}}$ agree with the
predictions of the Standard Model and allow a limit to be set
on graviton emission in one large extra dimension.
The limit obtained on the fundamental mass scale $M_D$ is 1.69 TeV/c$^2$
at 95$\%$ CL, with an expected limit of 1.71 TeV/c$^2$.
\end{shortabs}

\vfill

\begin{center}
\DpSubmit \ \\          
\DpComment \ \\
\DpEMail \ \\
\end{center}

\vfill
\clearpage

\headsep 10.0pt

\addtolength{\textheight}{10mm}
\addtolength{\footskip}{-5mm}
\begingroup
%
\newcommand{\DpName}[2]{\hbox{#1$^{\ref{#2}}$},\hfill}
\newcommand{\DpNameTwo}[3]{\hbox{#1$^{\ref{#2},\ref{#3}}$},\hfill}
\newcommand{\DpNameThree}[4]{\hbox{#1$^{\ref{#2},\ref{#3},\ref{#4}}$},\hfill}
\newskip\Bigfill \Bigfill = 0pt plus 1000fill
\newcommand{\DpNameLast}[2]{\hbox{#1$^{\ref{#2}}$}\hspace{\Bigfill}}

%
\footnotesize
\noindent
\DpName{J.Abdallah}{LPNHE}
\DpName{P.Abreu}{LIP}
\DpName{W.Adam}{VIENNA}
\DpName{P.Adzic}{DEMOKRITOS}
\DpName{T.Albrecht}{KARLSRUHE}
\DpName{R.Alemany-Fernandez}{CERN}
\DpName{T.Allmendinger}{KARLSRUHE}
\DpName{P.P.Allport}{LIVERPOOL}
\DpName{U.Amaldi}{MILANO2}
\DpName{N.Amapane}{TORINO}
% ...
\normalsize
\endgroup

\newpage
\titlefoot{Department of Physics and Astronomy, Iowa State
     University, Ames IA 50011-3160, USA
    \label{AMES}}
\titlefoot{Physics Department, Universiteit Antwerpen,
     Universiteitsplein 1, B-2610 Antwerpen, Belgium
    \label{ANTWERP}}
\titlefoot{IIHE, ULB-VUB,
     Pleinlaan 2, B-1050 Brussels, Belgium
    \label{BRUSSELS}}
\titlefoot{Physics Laboratory, University of Athens, Solonos Str.
     104, GR-10680 Athens, Greece
    \label{ATHENS}}
\titlefoot{Department of Physics, University of Bergen,
     All\'egaten 55, NO-5007 Bergen, Norway
    \label{BERGEN}}
\titlefoot{Dipartimento di Fisica, Universit\`a di Bologna and INFN,
     Viale C. Berti Pichat 6/2, IT-40127 Bologna, Italy
    \label{BOLOGNA}}
\titlefoot{Centro Brasileiro de Pesquisas F\'{\i}sicas, rua Xavier Sigaud 150,
     BR-22290 Rio de Janeiro, Brazil
    \label{BRASIL-CBPF}}
\titlefoot{Inst. de F\'{\i}sica, Univ. Estadual do Rio de Janeiro,
     rua S\~{a}o Francisco Xavier 524, Rio de Janeiro, Brazil
    \label{BRASIL-IFUERJ}}
\titlefoot{Coll\`ege de France, Lab. de Physique Corpusculaire, IN2P3-CNRS,
     FR-75231 Paris Cedex 05, France
    \label{CDF}}
\titlefoot{CERN, CH-1211 Geneva 23, Switzerland
    \label{CERN}}
\titlefoot{Institut de Recherches Subatomiques, IN2P3 - CNRS/ULP - BP20,
     FR-67037 Strasbourg Cedex, France
    \label{CRN}}
\titlefoot{Now at DESY-Zeuthen, Platanenallee 6, D-15735 Zeuthen, Germany
    \label{DESY}}
\titlefoot{Institute of Nuclear Physics, N.C.S.R. Demokritos,
     P.O. Box 60228, GR-15310 Athens, Greece
    \label{DEMOKRITOS}}
\titlefoot{FZU, Inst. of Phys. of the C.A.S. High Energy Physics Division,
     Na Slovance 2, CZ-182 21, Praha 8, Czech Republic
    \label{FZU}}
\titlefoot{Dipartimento di Fisica, Universit\`a di Genova and INFN,
     Via Dodecaneso 33, IT-16146 Genova, Italy
    \label{GENOVA}}
\titlefoot{Institut des Sciences Nucl\'eaires, IN2P3-CNRS, Universit\'e
     de Grenoble 1, FR-38026 Grenoble Cedex, France
    \label{GRENOBLE}}
\titlefoot{Helsinki Institute of Physics and Department of Physical Sciences,
     P.O. Box 64, FIN-00014 University of Helsinki, 
     \indent~~Finland
    \label{HELSINKI}}
\titlefoot{Joint Institute for Nuclear Research, Dubna, Head Post
     Office, P.O. Box 79, RU-101 000 Moscow, Russian Federation
    \label{JINR}}
\titlefoot{Institut f\"ur Experimentelle Kernphysik,
     Universit\"at Karlsruhe, Postfach 6980, DE-76128 Karlsruhe,
     Germany
    \label{KARLSRUHE}}
\titlefoot{Institute of Nuclear Physics PAN,Ul. Radzikowskiego 152,
     PL-31142 Krakow, Poland
    \label{KRAKOW1}}
\titlefoot{Faculty of Physics and Nuclear Techniques, University of Mining
     and Metallurgy, PL-30055 Krakow, Poland
    \label{KRAKOW2}}
\titlefoot{LAL, Univ Paris-Sud, CNRS/IN2P3, Orsay, France
    \label{LAL}}
\titlefoot{School of Physics and Chemistry, University of Lancaster,
     Lancaster LA1 4YB, UK
    \label{LANCASTER}}
\titlefoot{LIP, IST, FCUL - Av. Elias Garcia, 14-$1^{o}$,
     PT-1000 Lisboa Codex, Portugal
    \label{LIP}}
\titlefoot{Department of Physics, University of Liverpool, P.O.
     Box 147, Liverpool L69 3BX, UK
    \label{LIVERPOOL}}
\titlefoot{Dept. of Physics and Astronomy, Kelvin Building,
     University of Glasgow, Glasgow G12 8QQ, UK
    \label{GLASGOW}}
\titlefoot{LPNHE, IN2P3-CNRS, Univ.~Paris VI et VII, Tour 33 (RdC),
     4 place Jussieu, FR-75252 Paris Cedex 05, France
    \label{LPNHE}}
\titlefoot{Department of Physics, University of Lund,
     S\"olvegatan 14, SE-223 63 Lund, Sweden
    \label{LUND}}
\titlefoot{Universit\'e Claude Bernard de Lyon, IPNL, IN2P3-CNRS,
     FR-69622 Villeurbanne Cedex, France
    \label{LYON}}
\titlefoot{Dipartimento di Fisica, Universit\`a di Milano and INFN-MILANO,
     Via Celoria 16, IT-20133 Milan, Italy
    \label{MILANO}}
\titlefoot{Dipartimento di Fisica, Univ. di Milano-Bicocca and
     INFN-MILANO, Piazza della Scienza 3, IT-20126 Milan, Italy
    \label{MILANO2}}
\titlefoot{IPNP of MFF, Charles Univ., Areal MFF,
     V Holesovickach 2, CZ-180 00, Praha 8, Czech Republic
    \label{NC}}
\titlefoot{NIKHEF, Postbus 41882, NL-1009 DB
     Amsterdam, The Netherlands
    \label{NIKHEF}}
\titlefoot{National Technical University, Physics Department,
     Zografou Campus, GR-15773 Athens, Greece
    \label{NTU-ATHENS}}
\titlefoot{Physics Department, University of Oslo, Blindern,
     NO-0316 Oslo, Norway
    \label{OSLO}}
\titlefoot{Dpto. Fisica, Univ. Oviedo, Avda. Calvo Sotelo
     s/n, ES-33007 Oviedo, Spain
    \label{OVIEDO}}
\titlefoot{Department of Physics, University of Oxford,
     Keble Road, Oxford OX1 3RH, UK
    \label{OXFORD}}
\titlefoot{Dipartimento di Fisica, Universit\`a di Padova and
     INFN, Via Marzolo 8, IT-35131 Padua, Italy
    \label{PADOVA}}
\titlefoot{Rutherford Appleton Laboratory, Chilton, Didcot
     OX11 OQX, UK
    \label{RAL}}
\titlefoot{Dipartimento di Fisica, Universit\`a di Roma II and
     INFN, Tor Vergata, IT-00173 Rome, Italy
    \label{ROMA2}}
\titlefoot{Dipartimento di Fisica, Universit\`a di Roma III and
     INFN, Via della Vasca Navale 84, IT-00146 Rome, Italy
    \label{ROMA3}}
\titlefoot{DAPNIA/Service de Physique des Particules,
     CEA-Saclay, FR-91191 Gif-sur-Yvette Cedex, France
    \label{SACLAY}}
\titlefoot{Instituto de Fisica de Cantabria (CSIC-UC), Avda.
     los Castros s/n, ES-39006 Santander, Spain
    \label{SANTANDER}}
\titlefoot{Inst. for High Energy Physics, Serpukov
     P.O. Box 35, Protvino, (Moscow Region), Russian Federation
    \label{SERPUKHOV}}
\titlefoot{J. Stefan Institute, Jamova 39, SI-1000 Ljubljana, Slovenia
    \label{SLOVENIJA1}}
\titlefoot{Laboratory for Astroparticle Physics,
     University of Nova Gorica, Kostanjeviska 16a, SI-5000 Nova Gorica, Slovenia
    \label{SLOVENIJA2}}
\titlefoot{Department of Physics, University of Ljubljana,
     SI-1000 Ljubljana, Slovenia
    \label{SLOVENIJA3}}
\titlefoot{Fysikum, Stockholm University,
     Box 6730, SE-113 85 Stockholm, Sweden
    \label{STOCKHOLM}}
\titlefoot{Dipartimento di Fisica Sperimentale, Universit\`a di
     Torino and INFN, Via P. Giuria 1, IT-10125 Turin, Italy
    \label{TORINO}}
\titlefoot{INFN,Sezione di Torino and Dipartimento di Fisica Teorica,
     Universit\`a di Torino, Via Giuria 1,
     IT-10125 Turin, Italy
    \label{TORINOTH}}
\titlefoot{Dipartimento di Fisica, Universit\`a di Trieste and
     INFN, Via A. Valerio 2, IT-34127 Trieste, Italy
    \label{TRIESTE}}
\titlefoot{Istituto di Fisica, Universit\`a di Udine and INFN,
     IT-33100 Udine, Italy
    \label{UDINE}}
\titlefoot{Univ. Federal do Rio de Janeiro, C.P. 68528
     Cidade Univ., Ilha do Fund\~ao
     BR-21945-970 Rio de Janeiro, Brazil
    \label{UFRJ}}
\titlefoot{Department of Radiation Sciences, University of
     Uppsala, P.O. Box 535, SE-751 21 Uppsala, Sweden
    \label{UPPSALA}}
\titlefoot{IFIC, Valencia-CSIC, and D.F.A.M.N., U. de Valencia,
     Avda. Dr. Moliner 50, ES-46100 Burjassot (Valencia), Spain
    \label{VALENCIA}}
\titlefoot{Institut f\"ur Hochenergiephysik, \"Osterr. Akad.
     d. Wissensch., Nikolsdorfergasse 18, AT-1050 Vienna, Austria
    \label{VIENNA}}
\titlefoot{Inst. Nuclear Studies and University of Warsaw, Ul.
     Hoza 69, PL-00681 Warsaw, Poland
    \label{WARSZAWA}}
\titlefoot{Now at University of Warwick, Coventry CV4 7AL, UK
    \label{WARWICK}}
\titlefoot{Fachbereich Physik, University of Wuppertal, Postfach
     100 127, DE-42097 Wuppertal, Germany \\
\noindent
{$^\dagger$~deceased}
    \label{WUPPERTAL}}
\nopagebreak
\clearpage

\headsep 30.0pt
\end{titlepage}

%
\pagenumbering{arabic}                              
\setcounter{footnote}{0}                            %
\large
\section{Introduction}

The Standard Model (SM) has been thoroughly tested at the CERN LEP $e^+e^-$
collider~\cite{{lep},{lep1}}. No sign of statistically significant deviations 
from it or evidence for new physics phenomena beyond it have been found up to 
the highest LEP centre-of-mass energies of about 209 GeV. Yet the SM cannot be 
the final picture, because of several theoretical problems. One is known as 
the hierarchy problem and is related to the observed weakness of gravity in 
comparison with other interactions. This may be expressed by the observation 
that the reduced Planck mass, $M_{Pl} = \sqrt{1/G_N} \sim 2.4 \cdot 10^{15}$ 
TeV/c$^2$, where $G_N$ is Newton's coupling constant, is much larger than the 
0.1-1 TeV/c$^2$ scale of the electroweak symmetry breaking. 
\par
A step towards the solution of this puzzle was proposed in 1998 by Arkani-Hamed, 
Dimopoulos and Dvali (ADD)~\cite{add}, assuming the existence of large extra 
spatial dimensions (ED). Models with one ED were proposed a long time ago in 
connection with gravity and its unification with electromagnetism in the papers 
of Kaluza and Klein (KK)~\cite{{kaluza},{kaluza1},{kaluza2}}. More recently, 
with the appearance of string theory, the existence of several EDs was advocated, 
but their size was thought to be close to the Planck length, $R \sim 1/M_{Pl} \sim10^{-33}$ cm. 
In this case EDs would be completely out of the reach of present and planned 
colliders. The novel suggestion of ADD was the possible existence of large EDs 
with a fundamental Planck mass close to the electroweak scale, in fact implying 
that non-trivial physics ``ends'' at energies of about 1 TeV. In the ADD model 
all the SM particles are supposed to live on a 3D brane corresponding to our 
usual space, while gravitons are allowed to propagate into the bulk. Thus the 
weakness of gravity is simply due to its dilution in the volume of the EDs. 
\par  
Assuming flat EDs and compactification on a torus, Gauss' law gives: 

\begin{equation}
M^2_{Pl} = R^n M_D^{n+2},
\label{eq:gausslaw}
\end{equation}

\noindent where $R$ is the radius of the ED and $M_D$ is the fundamental Planck 
scale in the D-dimensional space-time (D=4+$n$). With $M_D \sim$1 TeV/c$^2$ and 
$n$=1, eq.~(\ref{eq:gausslaw}) implies a modification of Newton's law over solar 
system distances which is not observed. So the possibility that $n$=1 is usually 
considered to be falsified. On the other hand for $n\geq$2, $R<$1~mm and tests of 
gravity are only recently reaching these small distances~\cite{coyle}. For $n\geq$3, 
$R<$1~nm and no gravity test exists which can falsify the model. 
\par
The graviton, confined within flat EDs of size $R$, has a uniform spectrum of 
excitations, which, from the point of view of a 4D observer, will be seen as a 
KK tower of states, with masses uniformly spaced between $1/R$ $(\sim 10^{-32/n}$ 
TeV/c$^2$) and $M_D$. In particle collisions at accelerators and in the cosmos, 
gravitons can be emitted, but they escape immediately into the bulk, with momentum 
conservation in all the dimensions, and are therefore detectable via a missing 
energy signature. Each KK state is very weakly coupled, yet the number of states 
is very large, which turns into a sizable cross-section for graviton emission. 
Astrophysics yields strong constraints for $n$=2,3 based on observations of 
supernova SN1987A and on the behaviour of neutron stars~\cite{{astro},{astro1}}. 
The limits on the $M_D$ scale vary from 20 to 40 TeV/c$^2$ and 2 to 3 TeV/c$^2$, 
respectively, and seem to rule out the ADD model with $M_D$=1 TeV/c$^2$. They are 
however based on many assumptions with differences of a factor of 2-3 between 
different calculations. For larger $n$ they become much weaker. 
\par 
For $n\geq$2 limits on graviton emission have been obtained at the LEP collider~\cite{{al},{de},{l3},{op}} 
and at the Tevatron~\cite{{tevatron},{tevatron1}}. At LEP the direct graviton 
emission reaction \eeGg ($GZ$) has been studied: for $n\geq$2 the photon spectrum 
peaks at low energies and at small emission angles~\cite{grw}. No excess with 
respect to the SM predictions has been found and a combination of the LEP results 
yielded $M_D>$1.60 (0.80) TeV/c$^2$ for $n$=2 (6) at the 95$\%$ Confidence Level 
(CL)~\cite{wg}.  
\par 
Recently the ADD model has been reconsidered by Giudice, Plehn and Strumia (GPS)~\cite{gps}, 
who have focused on the infrared (IR) behaviour of the model in connection with 
limits at colliders versus gravity and astrophysics constraints. They considered a 
distorted version of the ADD model with the same properties in the ultraviolet (UV) 
region, but satisfying observational and astrophysical limits in the large distance 
regime. They showed that the introduction of an IR cut-off in the ADD model evades 
the constraints from astrophysics and gravity for small $n$, including $n$=1, given 
the energy resolution of the collider experiments. This IR cut-off is equivalent to 
a slight deformation or warping of the otherwise flat EDs. They started from the 
Randall and Sundrum type~1 model (RS1)~\cite{rs1} and considered the limit of slightly 
warped but large ED, resulting in a moderately large total warp factor. In RS1 the 
visible brane is located at $y$=0, where $y$ is the coordinate in the extra dimension, 
and the Planck brane at $y=\pi R$. The line element is non-factorisable due to the 
warping factor 
\begin{eqnarray}
ds^2 = e^{2 \sigma(y)} \eta _{\mu \nu} dx^{\mu}dx^{\nu} + dy^2 
\end{eqnarray}
\noindent with $\sigma(y) = \mu |y|$. Here $\mu$ is a mass parameter due to the warp, 
which has a value 50 MeV/c$^2$ $\leq \mu <<$ 1 TeV/c$^2$ and introduces an IR cut-off. 
This cut-off implies a mass of the graviton which is inaccessible for cosmological 
processes, but which has no significant implications for the high energy collider 
signal in the UV region of the KK spectra. In particular, the relation between the 
fundamental mass scale in 5 dimensions and the 4D Planck mass becomes
\begin{equation}
M^2_{Pl} = \frac{M^3_5}{2 \mu \pi} \left( e^{2 \mu R \pi} - 1 \right),
\end{equation}
where $R$ is the radius of the compactified ED. Hence the one ED can still be large, 
but unobserved as a modification of Newton's law or in the cosmological low energy 
processes. In this model the hierarchy between the Fermi and Planck scales is generated 
by two factors, the large ED and warping. It can be seen that for $\mu << R^{-1}$ the 
ADD limit, eq.~(\ref{eq:gausslaw}), is obtained.  
\par
Since a search for graviton emission with $n$=1 was not performed in the previous 
publication~\cite{de} and since the results cannot be inferred from the limits already 
given for $n\geq$2 because the photon energy spectra differ noticeably for different 
values of $n$~\cite{{grw},{gps}}, the DELPHI data were reanalysed and the results will 
be presented here. The paper is organized as follows: Section 2 recalls briefly the 
experimental details, the analysis is discussed in Section 3, Section 4 presents the 
results and the conclusions are given in Section 5.

\section{Detector and data preselection}

\par The general criteria for the selection of single-photon events are based mainly 
on the electromagnetic calorimeters and on the tracking system of the DELPHI 
detector~\cite{{delphi},{delphi1}}. All the three major electromagnetic calorimeters 
in DELPHI, the High density Projection Chamber (HPC), the Forward ElectroMagnetic 
Calorimeter (FEMC) and the Small angle TIle Calorimeter (STIC), have been used in 
the single-photon reconstruction. The STIC accepted photons at very small polar 
angle~\footnote{In the DELPHI coordinate system, the z axis is along the electron 
beam direction and the polar angle to the z axis is called $\theta$.}, the FEMC 
covered intermediate angles, and large angles with respect to the beams were covered 
by the HPC. Hermeticity Taggers were used to ensure detector hermeticity for additional 
neutral particles in the angular region around 45$^{\circ}$ between HPC and FEMC, not 
covered by the calorimeters. The DELPHI tracking system and the taggers were used as 
a veto. A detailed description of the trigger conditions and efficiencies of the 
calorimeters is given in a previous publication~\cite{de}, where the rejection of 
events in which charged particles were produced is also discussed. 
\par
The study was done with data taken during the 1997-2000 runs at $e^+e^-$ centre-of-mass 
energies from 180 to 209 GeV, corresponding to an integrated luminosity of $\sim$ 
650 $\mathrm {pb^{-1}}$, with the subdetectors relevant for the analysis all fully 
operational.
\par
The single-photon events were selected in two stages. In the first stage events with 
only one detected photon were preselected and compared to the SM process 
$e^+e^-~\rightarrow~\nu\bar{\nu}\gamma$. A likelihood ratio method was then used to 
maximize the sensitivity in the search for graviton production with $n$=1.
\par
Events with a photon in the HPC were selected by requiring a shower having a scaled 
energy $x_\gamma = E_\gamma/E_{beam} >$0.06, $\theta$ between 45$^\circ$ and 135$^\circ$, 
and no charged particle tracks. Photons in the FEMC were required to have a scaled 
energy $x_\gamma>$0.10 and a polar angle in the intervals 12$^\circ$~$< \theta <$~32$^\circ$ 
(148$^\circ$~$< \theta <$~168$^\circ$). Single photons in the STIC were preselected by 
requiring one shower with a scaled energy $x_\gamma>$0.30 and with 3.8$^\circ$~$< \theta <$~8$^\circ$ 
(172$^\circ$~$< \theta <$~176.2$^\circ$). Additional details about the preselection are 
given in~\cite{de}. In the single-photon event preselection events with more than one 
photon were accepted only if the other photons were at low angle ($\theta_\gamma$ $<$ 2.2$^\circ$), 
low energy ($E_\gamma<$0.8 GeV) or within 3$^\circ$, 15$^\circ$, 20$^\circ$ from the 
highest energy photon in the STIC, FEMC and HPC respectively. 

\section{Single-photon analysis}

The single-photon analysis has been discussed in detail in~\cite{de}, here we will 
recall the main points and underline the differences in the present analysis. 
\par 
Single-photon events can be faked by the QED reaction \eeeeg if the two electrons 
escape undetected along the beampipe or if the electrons are in the detector 
acceptance but are not detected by the experiment. This process has a very high 
cross-section, decreasing rapidly with increasing energy and polar angle of the 
photon. Its behaviour together with the rapid variation of efficiencies at low 
photon energy motivates the different calorimeter energy cuts in the preselection 
and additional energy-dependent cuts on the polar angle in the FEMC and STIC.
\par The remaining background from the \eeeeg process was 
calculated with the Monte Carlo program TEEG by D. Karlen~\cite{karlen} and two 
different event topologies were found to contribute, giving background at low 
and high photon energy respectively. Either both electrons were below the STIC 
acceptance or one of the electrons was in the DELPHI acceptance where it was 
wrongly identified as a photon, while the photon was lost for example in the 
gaps between the electromagnetic calorimeters not covered by the Hermeticity 
Taggers, or in masked crystals in the FEMC. 
\par The contribution from other processes has also been calculated: cosmic 
ray events, $\gamma\gamma$ collisions using PYTHIA 6.1~\cite{pythia} and 
BDK~\cite{{gammagamma},{gammagamma1}}, $e^+e^-~\rightarrow~\gamma\gamma(\gamma)$ 
according to Berends et al.~\cite{{berends1},{berends1a},{berends1b}},
$e^+e^-~\rightarrow~\mu\mu(\gamma)$ and $e^+e^-~\rightarrow~\tau\tau(\gamma)$ 
with KORALZ~\cite{{koralz},{koralz1}}, and four-fermion events with 
EXCALIBUR~\cite{berends2} and Grc4f~\cite{grc4f}.
\par The \eenngg~process was simulated by the 
KORALZ~\cite{{koralz},{koralz1}} program. A comparison of the cross-section 
predicted by KORALZ 4.02 with that predicted by NUNUGPV~\cite{{nunugpv},{nunugpv1}} 
and KK 4.19~\cite{kk2f} showed agreement at the percent level. This difference 
is negligible with respect to the statistical and systematic uncertainties in the 
present measurement. 
\par Simulated events for the irreducible contribution from $\nu\bar{\nu}\gamma$ 
production and other SM backgrounds were generated at the different centre-of-mass 
energies and passed through the full DELPHI simulation and reconstruction 
chain~\cite{{delphi},{delphi1}}.

\vspace{0.4cm}
\begin{center}
\begin{tabular}{|c|c|c|c|} \hline
   & N$_{observed}$ & N$_{e^+{e^-} \rightarrow \nu{\bar{\nu}}({\gamma)}}$ & 
     N$_{other~SM~background}$  \\ \hline
FEMC & 705 & 626$\pm$3 & 49.1 \\ \hline
HPC  & 498 & 540$\pm$4 & 0.6  \\ \hline
\end{tabular} \\
\vspace{0.3cm}
Table 1: The number of selected and expected single-photon events. 
\end{center}
\vspace{0.4cm}

\par
\begin{figure}[ht!]
\vspace*{-0.9cm}
\begin{center}
\mbox{\epsfig{figure=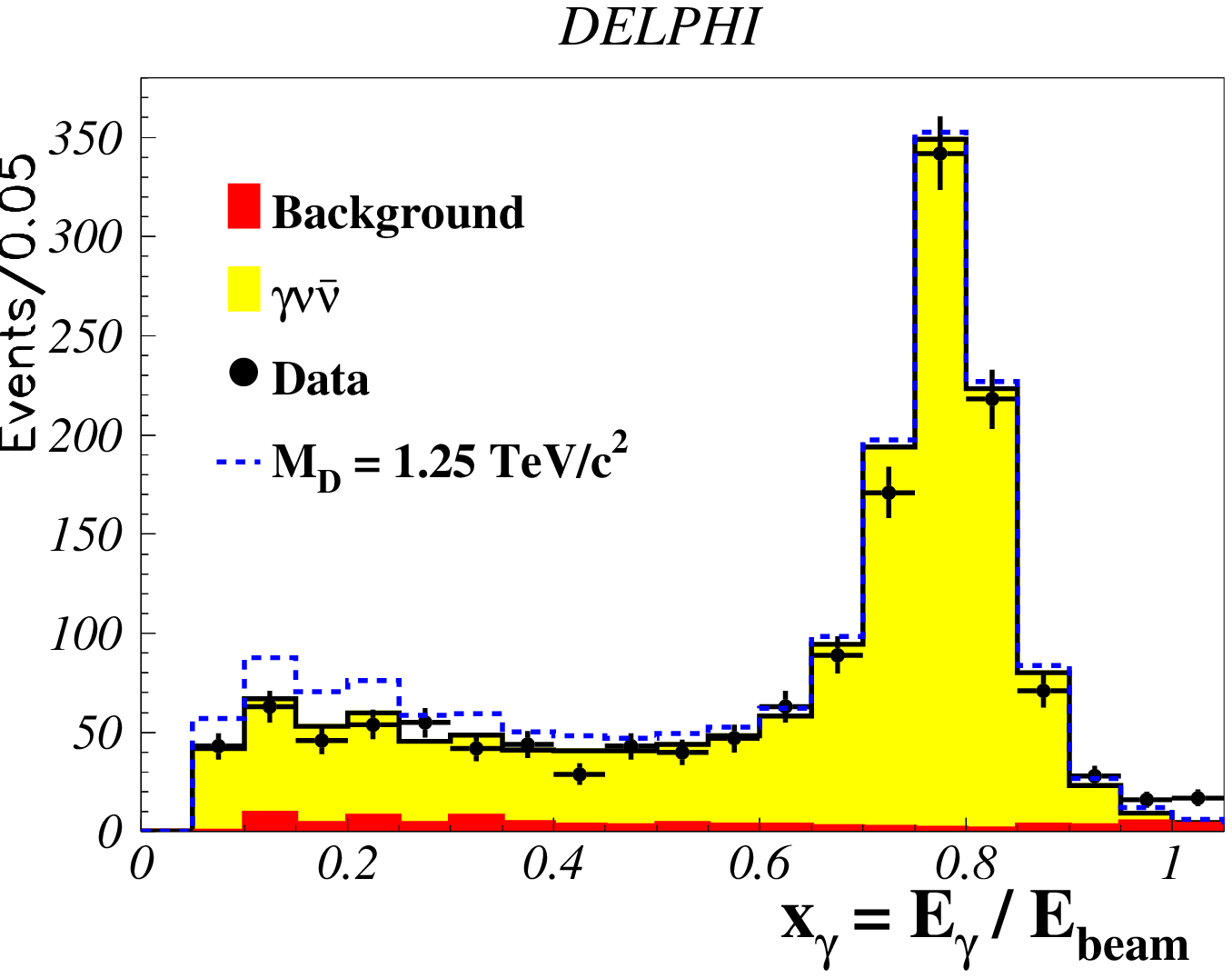,width=9.0cm}}
\end{center}
\caption[]{$x_\gamma$ of selected single photons.
The light shaded area is the expected distribution from \eenngg\ 
and the dark shaded area 
is the total background from other sources. Indicated in the plot is also 
the signal expected from \eeGg
for $n$=1 and $M_D$=1.25 TeV/c$^2$. 
}
\label{fig:xgamma}
\end{figure}

Figure~\ref{fig:xgamma} shows the $x_\gamma$ distribution of all preselected 
single-photon events. As discussed in the previous paper~\cite{de}, only single 
photon events in the HPC and FEMC were used for the subsequent analysis, since 
the $E_\gamma$ cuts in the STIC, needed to reduce the radiative Bhabha background, 
reject a large part of the ED signal even in the case $n$=1. 

\par Table 1 shows the total number of observed and expected events in the HPC 
and FEMC. The numbers are integrated over the LEP energies from 180 to 209 GeV 
and correspond to an overall luminosity of $\sim$650 $\mathrm {pb^{-1}}$.

\par  
A likelihood ratio method was used to select the final sample of single-photon 
events. This method allows the final selection to be optimised for excluding the 
cross section of a given signal assuming that no signal is present in the data 
sample. Hence the method optimises the background suppression for a given signal 
efficiency~\cite{anderson}. The likelihood ratio function used in this analysis 
is given by:
\begin{equation}
 {\cal L}_R = \frac{{\cal L}_S}{{\cal L}_B} = \frac{P_S(E_\gamma)}{P_B(E_\gamma)}  .
\end{equation}
The probability density functions ($P_{i=S,B}$) used to construct ${\cal L}_R$ were produced 
from the normalised photon energy distributions of the expected ED and SM background 
events, after passing through the same selection criteria. A low pass filter was also 
used to eliminate the high frequency statistical fluctuations from the final $P_i$ functions. 
An event was then selected as a candidate event if it passed the requirement 
${\cal L}_R$ $>$ ${\cal L}_R^{CUT}$. The value of ${\cal L}_R^{CUT}$ was determined by 
minimising the expected excluded cross section in the absence of a signal:
\begin{equation}
 \sigma^{min}({\cal L}_R^{CUT}) = {N_{95}^{min}({\cal L}_R^{CUT}) 
 \over \epsilon^{max}({\cal L}_R^{CUT}) \times L}  ,
 \label{eq:sigmamin}
\end{equation}
where $N_{95}^{min}$ is the upper limit on the number of signal events at 95$\%$ CL 
computed with the mono-channel version of the Bayesian method in~\cite{obraztsov}. 
$\epsilon^{max}$ is the efficiency for the signal and $L$ is the integrated luminosity. 
The values of $N_{95}^{min}$ and $\epsilon^{max}$ both decrease with an increasing value 
of ${\cal L}_R^{CUT}$. Their derivatives, however, behave differently which results in 
a well defined minimum of $\sigma^{min}({\cal L}_R^{CUT})$.

\par
The data collected at different centre-of-mass energies were analysed separately and 
different analyses were made depending on the electromagnetic calorimeter in which 
the photon was recorded. The ${\cal L}_R^{CUT}$ values obtained showed a variation of 
around $0.7 \pm 0.1$, though all the final selections contained a rejected region in 
the energy spectra that covered most of the $Z$-peak, as expected. In some cases the 
selection also implied a slightly stronger criteria for the overall minimum photon 
energy. Out of the preselected FEMC events, 262 passed the final selection with 250.6 
expected and from the HPC events, 255 were selected with 263.5 expected. The signal 
efficiency of the final selection was between 85\% and 90\% with respect to preselection 
level. The final experimental limit was obtained using a Bayesian multi-channel 
method~\cite{obraztsov} which combined the results of the 20 analyses, the data for the 
two calorimeters being grouped into 10 datasets between 180 and 209 GeV centre-of-mass 
energy. The method takes into account all the available information (such as the fraction 
of the signal and the average background in each subdetector and in each data subsample) 
in order to properly calculate the final limit.   

\section{Limit on the production of gravitons}

\par The differential cross-section for \eeGg\ has been calculated in~\cite{{grw},{gps}} 
and is given by: 

\begin{equation}
 {{d^2}\sigma \over d{x_\gamma}d\cos{\theta_\gamma}} = 
  {\alpha \over 32s} {\pi^{n \over 2} \over \Gamma({n \over 2})} 
  \left({\sqrt{s} \over M_D}\right)^{n+2} f({x_\gamma},\cos{\theta_\gamma}),  
\label{eq:dsdxdcost}
\end{equation}

\noindent with 

\begin{equation}
 f(x,y) = {2(1-x)^{{n \over 2}-1} \over x(1-y^2)}[(2-x)^2(1-x+x^2) \\
  - 3{y^2}{x^2}(1-x) - {y^4}{x^4}].
\label{eq:fxy}
\end{equation}

Initial state radiation can produce additional photons that would cause a signal 
event to be rejected in a single-photon analysis. The expected signal cross-section 
has therefore been corrected with a radiator approximation method~\cite{nicrosini}. 
\par
\begin{figure}[ht!]
\vspace*{-0.8cm}
\begin{center}
\mbox{\epsfig{figure=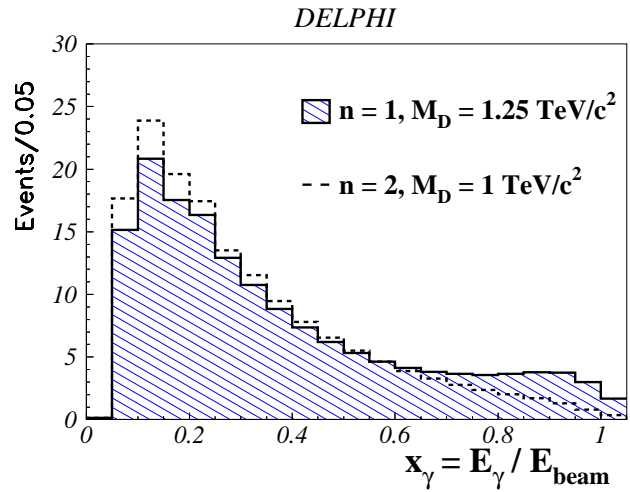,width=9.0cm}}
\end{center}
\caption[]{$x_\gamma$ of expected single photons in the HPC and FEMC 
from \eeGg
with $n$=1, $M_D$=1.25 TeV/c$^2$ and $n$=2, $M_D$=1 TeV/c$^2$, 
corrected for calorimeter efficiency and resolution. 
MC expectations are normalized to the luminosity of the 
combined data set in Fig. 1.
}
\label{fig:esmeared}
\end{figure}

 For $n>$1 the differential distribution, eq.~(\ref{eq:fxy}), is peaked at 
small $E_\gamma$ and $\theta_\gamma$, for $n$=1 instead a singularity is present 
at $x_\gamma$=1, which makes the distribution qualitatively different from the 
others. For instance the ratio of the cross-sections, eq.~(\ref{eq:dsdxdcost}) and 
eq.~(\ref{eq:fxy}), for $n$=1 and $n$=2 is independent of $\theta_\gamma$, and increases 
from $\sim$1.5 at small $x_\gamma$ to $\sim$22 at $x_\gamma$=0.995 for $M_D$=1 TeV/c$^2$ 
and $\sqrt{s}$=208 GeV. In order to take into account detector effects, the theoretical 
ED cross-section has been corrected for efficiency and energy resolution in the 
calorimeters, using a parameterization developed in the $\nu\bar{\nu}\gamma$ analysis. 
The theoretical energy distributions for $n$=1 and 2 smeared in the HPC and FEMC are 
shown in Fig.~\ref{fig:esmeared}. 
\par As can be seen in Fig.~\ref{fig:xgamma}, the single photon data measured by 
DELPHI were well compatible with expectations from SM processes and no evidence for 
graviton production was found. 
\par All DELPHI data with $\sqrt{s}>$180 GeV were used and a dedicated selection for 
each bin in $\sqrt{s}$ was made as described in the previous section. These results 
were combined to give a 95$\%$ CL cross-section limit for one extra dimension of 
0.171 $\mathrm {pb}$ at 208 GeV, with an expected limit of 0.166 $\mathrm {pb}$. In 
terms of the parameter $p = (1/M_{D})^3$, to which the $n$=1 signal cross-section is 
proportional, the combined log-likelihood function of the Bayesian formula was 
practically parabolic. $p$ is estimated to be ($0.009 \pm 0.098$) (TeV/c$^2$)$^{-3}$ 
and is therefore consistent with zero. The obtained limit on the fundamental mass 
scale is $M_D>$1.69 TeV/c$^2$ at 95$\%$ CL (with 1.71 TeV/c$^2$ expected limit) in the 
$n$=1 analysis. As a comparison, the cross-section limits in the previous analysis 
for $n$=2-6 varied between 0.14 and 0.18 $\mathrm {pb}$, and the obtained limits for 
$M_D$ between 1.31 TeV/c$^2$ ($n$=2) and 0.58 TeV/c$^2$ ($n$=6). Since the characteristic 
peak of the $n=1$ photon spectrum at $x_\gamma = 1$ is less prominent after including 
detector effects, the cross section limit is similar to those obtained for $n > 1$. 
The same systematic errors were considered as in the previous analysis~\cite{de}, namely 
trigger and identification efficiency, calorimeter energy scale and background, and the 
effect on the $M_D$ limit from the systematic errors in the $n$=1 analysis was estimated 
to be less than 4\%.

\section{Conclusions}

We have re-analysed single-photon events detected with DELPHI at LEP2 during 1997-2000 
at centre-of-mass energies between 180 and 209 GeV to study graviton production with 
$n$=1 large extra dimensions, motivated by the model of Giudice, Plehn and Strumia~\cite{gps}. 
Since the measured single-photon cross-sections are in agreement with the expectations 
from the SM process \eenngg, the absence of an excess of events has been used to set a 
limit of 1.69 TeV/c$^2$ at 95$\%$ CL on the fundamental mass scale for $n$=1 ED.

\newpage
\subsection*{Acknowledgements}
\vskip 3 mm
We are greatly indebted to our technical 
collaborators, to the members of the CERN-SL Division for the excellent 
performance of the LEP collider, and to the funding agencies for their
support in building and operating the DELPHI detector.\\
We acknowledge in particular the support of \\
Austrian Federal Ministry of Education, Science and Culture,
GZ 616.364/2-III/2a/98, \\
FNRS--FWO, Flanders Institute to encourage scientific and technological 
research in the industry (IWT) and Belgian Federal Office for Scientific, 
Technical and Cultural affairs (OSTC), Belgium, \\
FINEP, CNPq, CAPES, FUJB and FAPERJ, Brazil, \\
Ministry of Education of the Czech Republic, project LC527, \\
Academy of Sciences of the Czech Republic, project AV0Z10100502, \\
Commission of the European Communities (DG XII), \\
Direction des Sciences de la Mati$\grave{\mbox{\rm e}}$re, CEA, France, \\
Bundesministerium f$\ddot{\mbox{\rm u}}$r Bildung, Wissenschaft, Forschung 
und Technologie, Germany,\\
General Secretariat for Research and Technology, Greece, \\
National Science Foundation (NWO) and Foundation for Research on Matter (FOM),
The Netherlands, \\
Norwegian Research Council,  \\
State Committee for Scientific Research, Poland, SPUB-M/CERN/PO3/DZ296/2000,
SPUB-M/CERN/PO3/DZ297/2000, 2P03B 104 19 and 2P03B 69 23(2002-2004),\\
FCT - Funda\c{c}\~ao para a Ci\^encia e Tecnologia, Portugal, \\
Vedecka grantova agentura MS SR, Slovakia, Nr. 95/5195/134, \\
Ministry of Science and Technology of the Republic of Slovenia, \\
CICYT, Spain, AEN99-0950 and AEN99-0761,  \\
The Swedish Research Council,      \\
The Science and Technology Facilities Council, UK, \\
Department of Energy, USA, DE-FG02-01ER41155, \\
EEC RTN contract HPRN-CT-00292-2002. \\


\clearpage

\end{document}